\documentclass[aps,prl,letterpaper,twocolumn,superscriptaddress]{revtex4-1}
\pdfoutput=1 

\usepackage{graphicx}
\usepackage{amssymb}
\usepackage{amsmath}
\usepackage{mathrsfs}
\usepackage{placeins}
\usepackage{natbib}
\usepackage[usenames]{color}
\usepackage[normalem]{ulem}
\renewcommand{\sout}[1]{}

\newcommand{\mbf}[1]{\boldsymbol{#1}}

  \newcommand{\Rey}{\mathrm{Re}}

  \newcommand{\Wi}{\mathrm{Wi}}

\newlength{\figwidth}
\newlength{\SCwidth}
\def\XXint#1#2#3{{\setbox0=\hbox{$#1{#2#3}{\int}$}
     \vcenter{\hbox{$#2#3$}}\kern-.5\wd0}}

\newcommand{\ASrevise}[1]{\textcolor{black}{#1}}
\newcommand{\MDGrevise}[1]{\textcolor{black}{#1}}
\newcommand{\RMMrevise}[1]{\textcolor{black}{#1}}

\newcommand{\lam}{\mathrm{lam}}

\begin{document}

\title{Critical-layer structures and mechanisms in elastoinertial turbulence}


\author{Ashwin Shekar}
\affiliation{Department of Chemical and Biological Engineering, University of Wisconsin-Madison, Madison WI 53706, USA}

\author{Ryan M. McMullen}
\affiliation{Graduate Aerospace Laboratories, California Institute of Technology, Pasadena CA 91125, USA}

\author{Sung-Ning Wang}
\affiliation{Department of Chemical and Biological Engineering, University of Wisconsin-Madison, Madison WI 53706, USA}

\author{Beverley J. McKeon}
\affiliation{Graduate Aerospace Laboratories, California Institute of Technology, Pasadena CA 91125, USA}

\author{Michael D. Graham}
\email{mdgraham@wisc.edu}
\affiliation{Department of Chemical and Biological Engineering, University of Wisconsin-Madison, Madison WI 53706, USA}

\date{\today}


\begin{abstract}

Simulations of elastoinertial turbulence (EIT) of a polymer solution
 at low Reynolds number 
are shown to display localized polymer stretch fluctuations. 
These are very similar to structures \MDGrevise{arising} from linear stability (Tollmien-Schlichting (TS) modes) and resolvent analyses: i.e., critical-layer structures localized where the mean fluid velocity equals the wavespeed.
Computation of self-sustained nonlinear TS waves reveals that the critical layer exhibits stagnation points that generate sheets of large polymer stretch. These kinematics may be the genesis of similar structures in EIT.

\end{abstract}


\maketitle


\emph{Turbulent drag reduction} is \MDGrevise{an} important and puzzling phenomen\MDGrevise{on} in the non-Newtonian flow of complex fluids.  Addition of \sout{small amounts of} polymers or micelle-forming surfactants to a liquid can lead to dramatic reductions in energy dissipation during turbulent flow while having a negligible effect on laminar flow\cite{Virk:1970uo}. 

In Newtonian channel or pipe flow, transition to turbulence occurs by a \MDGrevise{so-called} subcritical or ``bypass'' transition mechanism \MDGrevise{as flow rate, measured nondimensionally by Reynolds number, $\Rey${,} increases}: turbulence is initiated by finite-amplitude perturbations to the laminar flow profile, while the laminar flow remains linearly stable. 
While channel flow exhibits a two-dimensional linear instability leading to so-called Tollmien-Schlichting (TS) waves, the critical Reynolds number {$\Rey=5772$} 
is much higher than that observed for transition, so these are not traditionally viewed as playing an important role in Newtonian transition. 

For flowing polymer solutions under some conditions \MDGrevise{(low concentration, short polymer relaxation times),} 
transition to turbulence occurs via  the usual bypass transition.\sout{ from laminar to turbulent flow as flow rate, measured nondimensionally by Reynolds number, $\Rey$, increases.} With further increase in $\Rey$, \sout{onset of} drag reduction \sout{occurs}\MDGrevise{sets in,} and the flow eventual\RMMrevise{ly} approaches the so-called maximum drag reduction (MDR) asymptote, an \sout{experimental} upper bound on the degree of drag reduction that is {insensitive to} the details of the fluid.

Under other conditions, 
 \sout{however,} flow transitions directly from laminar flow into the MDR regime, and can do so at a Reynolds number where the flow would remain laminar if Newtonian \cite{Forame:1972ti,Hoyt:1977vj,Choueiri:2018it,Chandra:2018bq}. Recent experiments and simulations \cite{Samanta:2013el,Dubief:2013hh,Sid:2018gh} suggest that turbulence in this regime has structure very different from Newtonian, denoting it as ``elastoinertial turbulence'' (EIT).  Choueiri \emph{et al.} \cite{Choueiri:2018it} experimentally observed that at \sout{very low (i.e.~transitional)}\MDGrevise{transitional} Reynolds numbers and increasing polymer concentration, turbulence is first suppressed, leading to relaminarization,  and then reinitiated with an EIT structure and a level of drag corresponding to MDR. \sout{This result indicates that} \MDGrevise{Therefore,} there are actually two distinct types of turbulence in polymer solutions, one that is suppressed by viscoelasticity, and one that is promoted. 

 The present work reports computations and analysis that elucidate the mechanisms underlying \sout{elastoinertial turbulence}\MDGrevise{EIT}. We show that  EIT at low $\Rey$ has highly localized \ASrevise{polymer} stress fluctuations. Surprisingly, these strongly resemble linear Tollmien-Schlichting modes as well as the most strongly amplified fluctuations from the laminar state. Furthermore, the kinematics of self-sustained nonlinear TS waves generate sheetlike structures in the stress field \sout{ that are} similar to those observed in EIT.
 {The resemblance of structures at EIT to these  Newtonian phenomena may shed light on the observed near-universality of the MDR regime with regard to polymer properties.}

\paragraph{Formulation}

We consider \MDGrevise{pressure-driven channel flow}
with constant mass flux. The $x$, $y$ and $z$ axes are aligned with the streamwise \ASrevise{(overall flow)}, wall-normal and spanwise directions, respectively. 
Lengths are scaled by the half channel height $l$ so the dimensionless channel height $L_y=2$. The domain is periodic in $x$ and $z$ with periods $L_x$ and $L_z$. 
Velocity $\mbf{v}$ is scaled with the Newtonian laminar centerline velocity $U$\sout{ for the given mass flux}; time $t$ with $l/U$, and pressure $p$ with $\rho U^2$, where $\rho$ is the fluid density. The polymer stress tensor $\mbf{\tau}_p$ is related to the polymer conformation tensor $\mbf{\alpha}$ \MDGrevise{(second moment of the probability distribution for the polymer end-to-end vector)} through the FENE-P constitutive relation{, which models each polymer molecule as a pair of beads connected by a \sout{finitely extensible} nonlinear spring with maximum extensibility $b$}.
\sout{The maximum extensibility of the polymer molecules is $b$.} We solve the \MDGrevise{momentum, continuity and FENE-P equations:}
\sout{nondimensionalized Navier-Stokes equation coupled with the FENE-P constitutive equation:}
\begin{gather}
        \label{Eq_ns_momentum}
                \frac{\partial \mbf{v}}{\partial t} +
                \mbf{v} \cdot \mbf{\nabla v} = -
                \mbf{\nabla}p + \frac{\beta}{\Rey} \nabla^{2}\mbf{v} +
                \frac{\left(1 -\beta\right)}{\Rey \Wi_{}}\left(\mbf{\nabla} \cdot
                \mbf{\tau}_{\mathrm{p}}\right),
        \\ 
        \label{Eq_ns_continuity}
                  \mbf{\nabla} \cdot \mbf{v} = 0,
        \\                   
        \label{Eq_fenep_stress}
        	        \mbf{\tau}_p = \frac{\mbf{\alpha}}{1-\frac{\mathrm{tr}(\mbf{\alpha})}{b}} - \mbf{I},             
        \\                
        \label{Eq_fenep_conformation}
	    			\frac{\partial \mbf{\alpha}}{\partial t} +         
        			\mbf{v} \cdot \mbf{\nabla \alpha} -
        			\mbf{\alpha} \cdot \mbf{\nabla v} - 
        			\left( \mbf{\alpha} \cdot \mbf{\nabla v} \right)^{\mathrm{T}}
			    = \frac{-1}{\Wi_{}} \mbf{\tau}_p.
\end{gather}
\sout{The Reynolds number is}\MDGrevise{Here} $\Rey = \rho U l / (\eta_{\mathrm{s}} + \eta_{\mathrm{p}})$, where \sout{$(\eta_{\mathrm{s}} + \eta_{\mathrm{p}})$ is the total zero-shear rate viscosity (subscripts ``s'' and ``p" indicate solvent and polymer contributions, respectively).}
\MDGrevise{$\eta_s$ and $\eta_p$ are the solvent and polymer contributions to the zero-shear rate viscosity.}
The viscosity ratio $\beta = \eta_{\mathrm{s}} / (\eta_{\mathrm{s}} + \eta_{\mathrm{p}})$\MDGrevise{; polymer concentration is proportional to $1-\beta$}.  We fix $\beta=0.97$ and $b=6400$.
The Weissenberg number $\Wi = \lambda U/l$, where $\lambda$ is the polymer relaxation time, \MDGrevise{ measures the ratio between the relaxation time for the polymer and the shear time scale for the flow.} \MDGrevise{Below we report values of friction factor $f = \frac{2\tau_{w}}{\rho U^2}$, where $\tau_{w}$ is time- and area -averaged wall shear stress. This is a nondimensional measure of pressure drop or drag. Its value in laminar flow is denoted $f_\lam$.}

For \MDGrevise{the nonlinear} direct {numerical} simulations {(DNS)} \MDGrevise{described below}, a finite difference scheme and a fractional time step method are adopted for integrating the Navier-Stokes equation. Second-order Adams-Bashforth and Crank-Nicolson methods are used for convection and diffusion terms, respectively. The FENE-P equation is discretized using a high resolution central difference scheme \citep{kurganov2000new,Vaithianathan:2006dy, Dallas:2010gu}. No artificial diffusion is applied.
For the three-dimensional \MDGrevise{(3D)} simulations, \sout{the domain has dimensions} $(L_x,L_y,L_z)=(10,2,5)$; these were chosen to match \cite{Samanta:2013el}.  Typical resolution for the 3D runs at EIT is $(N_{x},N_{y},N_{z}) = (189,150,189)$. For the 2D runs at $\Rey = 3000$, \sout{wall normal resolution of} $N_{y} = 302$ is used.
For the \MDGrevise{linear analyses,}
Eqs.~\ref{Eq_ns_continuity}-\ref{Eq_fenep_conformation}, linearized around the laminar solution and Fourier-transformed in $x$ $z$, and $t$, are discretized in $y$ with a Chebyshev pseudospectral method. 
Typically, about 200 Chebyshev polynomials are sufficient for the resolvent calculations, whereas as many as 400 are required for \sout{satisfactory convergence of} the TS eigenmode. {The norm used in the resolvent calculations is the sum of the kinetic energy and a measure of the conformation tensor perturbation magnitude that is consistent with the non-Euclidean geometry of positive-definite tensors~\cite{hameduddin2019}.}

\paragraph{Nonlinear simulation results:}
\label{sec:nonlin}

\begin{figure}
 	\includegraphics[scale=0.4]{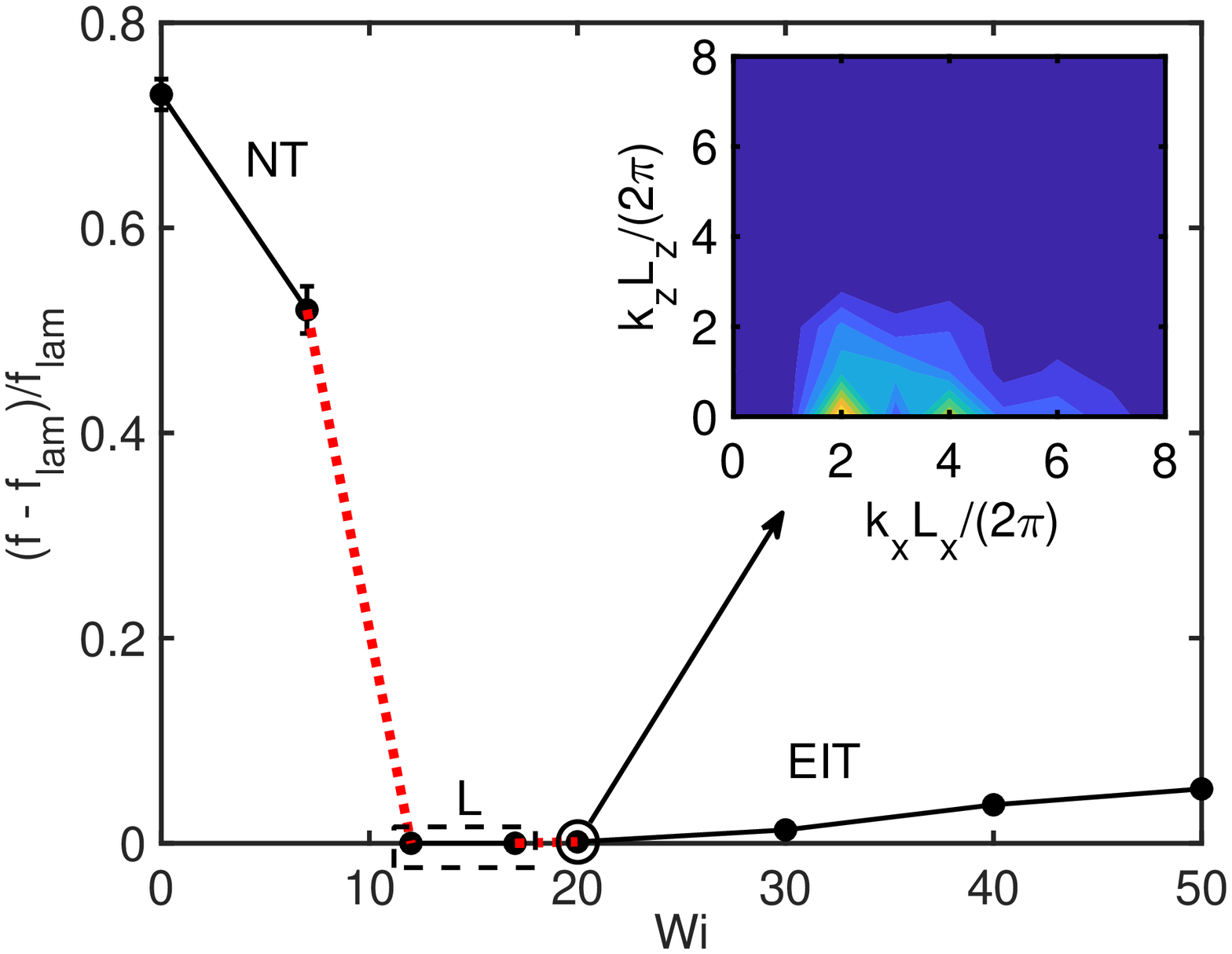}
	\caption[]{
    \sout{Normalized deviation of friction factor from laminar}\MDGrevise{Scaled friction factor} vs.~$\Wi$ at $\Rey=1500$. Abbreviations `NT', `L' and `EIT' stand for Newtonian-like turbulence, laminar and elastoinertial turbulence, respectively. In most cases, the error bars are smaller than the symbols. 
    Red dotted lines indicate the intervals of $\Wi$ in which the NT solution loses existence and the EIT solution comes into existence, respectively, as $\Wi$ increases.
    Inset shows the spatial spectrum of the wall normal velocity at $y=0$ for $\Wi = 20$. Here, $x$- and $z$-wavenumbers $k_x$ and $k_z$ are reported in scaled form, as $k_x L_x/2\pi$ and $k_z L_z/2\pi$
    . For inset, low is blue, high is yellow.}
 	\label{fig:fminflam}
\end{figure}

Fig.~\ref{fig:fminflam} illustrates 3D DNS results \MDGrevise{for scaled friction factor 
 $(f-f_\lam)/f_\lam$}
vs.~Weissenberg number $\Wi$ at $\Rey=1500$. 
At low but increasing $\Wi$, the flow is turbulent, with $f$ decreasing, indicating that the drag is reduced from the Newtonian value. In this regime{, which we denote NT,} the turbulence displays a streamwise vortex structure typical of Newtonian turbulence. With a further increase in $\Wi$, however, $f-f_\lam$ drops to zero -- the flow relaminarizes, as the NT regime loses existence. (At this $\Rey$ and all $\Wi$ considered here, the laminar state is linearly stable.) At still higher $\Wi$\sout{ however}, the flow, if seeded with a sufficiently energetic initial condition, becomes turbulent again, \sout{but} with a very low value of $f-f_\lam$ (consistent with experimental observations of~\cite{Choueiri:2018it} in pipe flow) and a very different structure: i.e. a new kind of turbulence comes into existence.
In this regime the flow structure corresponds to EIT as described by \cite{Samanta:2013el,Sid:2018gh}; we further analyze this structure below. 
\sout{These results indicate that}\MDGrevise{In short,} as $\Wi$ increases \MDGrevise{ from zero}, the self-sustaining mechanism of Newtonian turbulence is weakened by viscoelasticity, resulting in loss of existence of the NT state. \sout{On the other hand,}\MDGrevise{As} $\Wi$ increases further, a new nonlinear self-sustaining (i.e.~bypass transition) mechanism comes into play, resulting in EIT.

\begin{figure}
 	\includegraphics[width=\columnwidth]{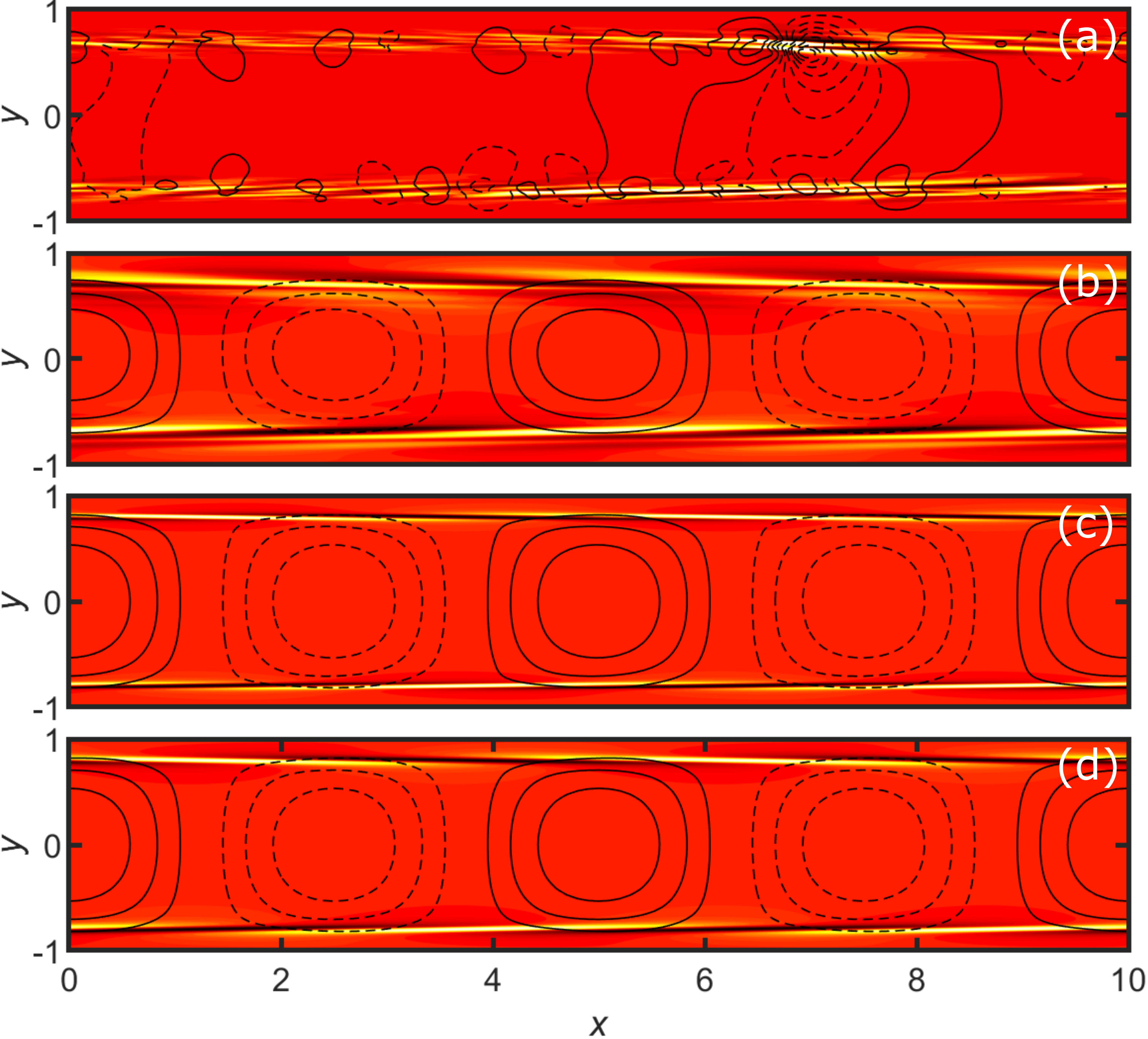}
	\caption[]{
	(a)~Snapshot of $v'$ (line contours) and $\alpha_{xx}'$ (filled contours) from 3D nonlinear DNS at $\Rey = 1500$, $\Wi=20$, where $'$ denotes fluctuations.
    (b)~Phase-matched average $(k_x L_x/2\pi,k_z L_z/2\pi)=(2,0)$ structures from 3D DNS. 
    (c)~Structure of the TS mode at $\Rey=1500, \Wi=20$, and the same wavenumbers as in (b). 
    (d)~Structure of the most strongly amplified resolvent mode at $\Rey=1500, \Wi=20$, the same wavenumbers as in (b), and $c = 0.37$.  In all plots, contour levels are symmetric about zero. For $v'$ dashed - negative, solid - positive.
    For $\alpha_{xx}'$ black - negative, red - zero and yellow - positive.
    }
 	\label{fig:Cxx_comparison}
\end{figure}

We now focus on the flow structure in the EIT regime. The inset in Fig.~\ref{fig:fminflam} shows a spatial spectrum of the wall normal velocity at $y=0$ (the channel centerplane), i.e., $|v(k_x,0,k_z)|$. The centerplane is chosen because it yields the cleanest spectra.
In the EIT regime, there is very strong spectral content when $k_z=0$, indicating the importance of \sout{two-dimensional}\MDGrevise{2D} mechanisms in the dynamics. Indeed,~\cite{Sid:2018gh} reports that EIT can arise in 2D simulations. Figure \ref{fig:Cxx_comparison}a shows a slice at {$z=2.5$} of the fluctuating wall normal velocity, $v'$, and fluctuating $xx$-component of the polymer conformation tensor, $\alpha'_{xx}$. Observe that {$\alpha'_{xx}$} is strongly localized near $y=\pm 0.7-0.8$. While tilted sheets of polymer stretch fluctuations have already been noted as characteristic of EIT \cite{Samanta:2013el}, the strong localization has not been previously observed, \sout{most likely}\MDGrevise{perhaps} because prior results have been at higher $\Rey$ and $\Wi$, i.e.~further from the point at which EIT comes into existence.  
  Fig.~\ref{fig:Cxx_comparison}b shows the dominant $(k_x L_x/2\pi,k_z L_z/2\pi)=(2,0)$ component of the $\Wi=20$ \sout{simulation} results, phase-matched and averaged over many snapshots. {Results for higher $k_x$ are very similar, exhibiting strong localization of stress fluctuations in the same narrow bands, as well as velocity fluctuations that span the channel height. }

\paragraph{Linear analyses:} 
\label{sec:lin}

\begin{figure}[t]
 	\includegraphics[scale=0.55]{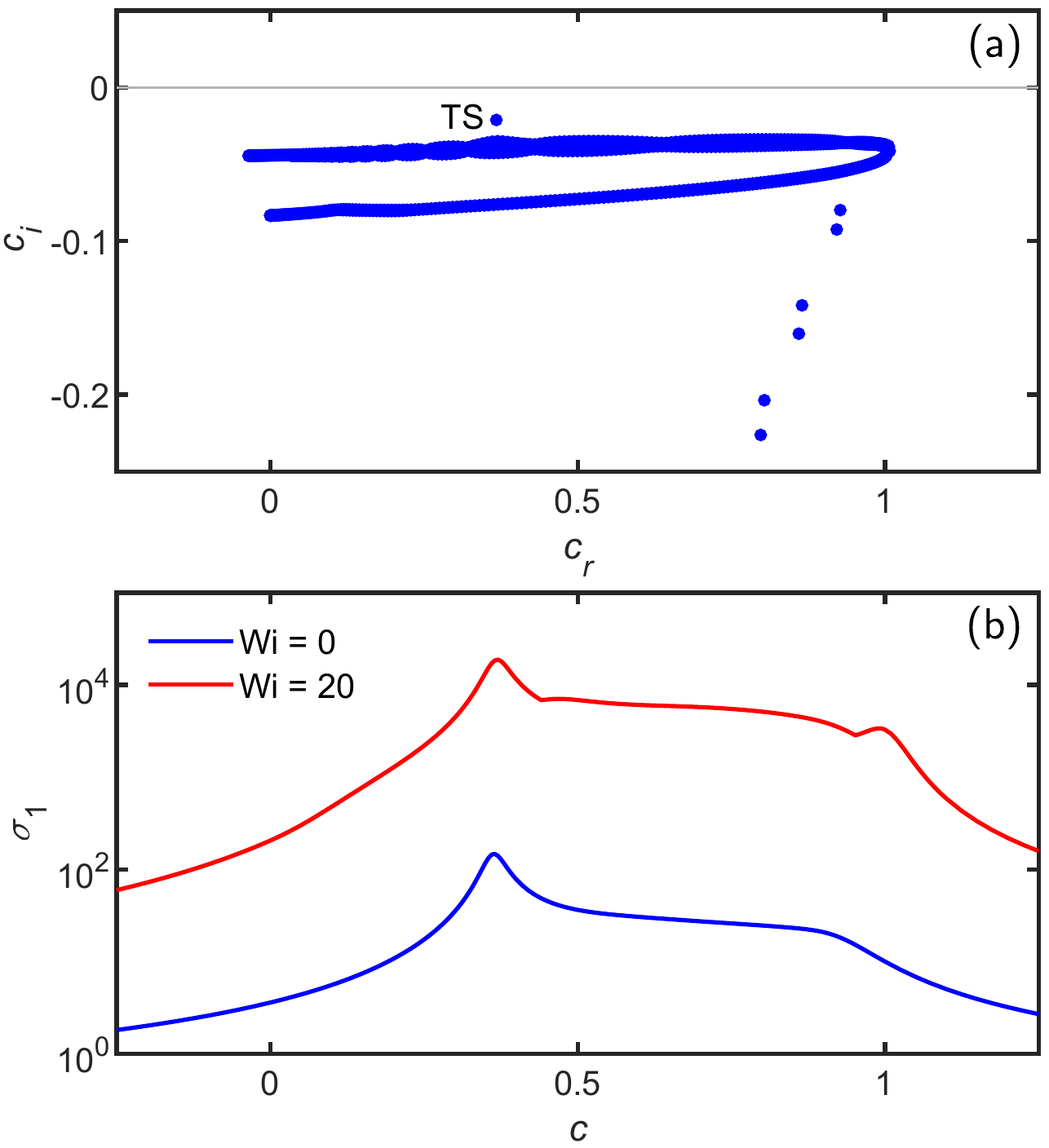}
	\caption[]{
    Eigenvalue spectrum for $(k_x L_x/2\pi,k_z L_z/2\pi)=(2,0)$ 
    with $\Wi=20$ and $\Rey=1500$. The eigenvalue labeled `TS' corresponds to the TS mode. (b) Leading singular value of the resolvent operator for $\Wi=0$ and $\Wi=20$, plotted on a logarithmic scale.
    }
    \label{fig:Combine_LA}
\end{figure}


To shed light on the origin of the highly localized \sout{region of }large stress fluctuations, we now consider the \sout{linearized} evolution of \MDGrevise{infinitesimal} perturbations \RMMrevise{to the laminar state} \MDGrevise{with given wavenumbers $k_x,k_z$}\sout{away from the laminar flow state}. Two approaches are used. \MDGrevise{The first is classical linear stability analysis, in which solutions of the form} \MDGrevise{$\phi(y)\exp\left[i\!\left(k_x x+k_z z-k_x ct\right)\right]$} \MDGrevise{are sought, resulting in an eigenvalue problem for the complex wavespeed} \RMMrevise{$c$}\sout{$c=c_r+i c_i$}. If \RMMrevise{any} $c_i>0$, {then the laminar state is linearly unstable -- }infinitesimal perturbations will grow exponentially. \sout{Otherwise, the flow is linearly stable.}\MDGrevise{If all $c_i<0$, the flow is linearly stable.}  The second approach is to determine the \sout{amplification}\MDGrevise{linear response} of \MDGrevise{ the laminar flow to} external forcing with\sout{a} given \sout{(}real\sout{)} frequency $\omega$ \RMMrevise{using the resolvent operator (frequency-space transfer function) of the linearized equations \cite{Schmid07,mckeon2010}}\sout{. Mathematically, this is done via the singular value decomposition of the resolvent operator (frequency-space transfer function) of the linearized equations}. In both analyses, the \sout{classical }concept of \emph{critical layers}\RMMrevise{, i.e., wall-normal positions where the fluid velocity equals the wavespeed of an eigenmode or resolvent mode,} is important\sout{ -- these are regions localized at wall-normal positions where the  fluid velocity equals the wavespeed of an eigenmode or singular mode}. While some recent studies suggest the importance of critical-layer mechanisms in viscoelastic shear flows \cite{Page:2015es,Lee:2017fb,Haward:2018hs,hameduddin2019}, they do not make as direct a connection to EIT as we \sout{will }illustrate here. 

Figure \ref{fig:Combine_LA}a shows the result of linear stability analysis (\MDGrevise{the \sout{spectrum of }eigenvalues $c$}\sout{the eigenvalue spectrum in terms of the complex wavespeed $c$})  for $\Wi=20$, $k_x L_x/2\pi=2$, $k_z=0$, the wavenumber corresponding to the dominant structures observed in the nonlinear simulations. All eigenvalues \RMMrevise{have $c_i<0$}\sout{fall in the lower half of the complex plane} -- the laminar flow is linearly stable.

Of note is the \sout{discrete }mode labeled `TS', the viscoelastic continuation of the classical Tollmien-Schlichting mode \cite{DrazinReid}. 
Viscoelasticity has only a weak effect on the TS eigenvalue, which changes from $c=0.362-0.019i$ to $c=0.368-0.022i$ between $\Wi=0$ and $\Wi=20$ \cite{zhang2013}. 
Despite the small change in $c$, the conformation tensor disturbance depends very strongly on $\Wi$; the peak value of $\alpha'_{xx}$ grows from zero at $\Wi=0$ to \sout{five orders of magnitude larger than}\MDGrevise{\sout{about}}\RMMrevise{$\sim\!10^5$} \MDGrevise{times} the peak value of $u'$ at $\Wi=20$. 

The structure of this eigenmode is shown for $\Wi=20$ in~Fig.~\ref{fig:Cxx_comparison}c. In the Newtonian case, the disturbance velocity field {is} 
a train of spanwise-oriented vortices that span the entire channel{;} this structure is only weakly modified \MDGrevise{even at high $\Wi$}\sout{ by even very strong viscoelasticity}. The polymer stress disturbance behaves very differently: at $\Wi=20$ it consists of highly inclined sheets that are extremely localized around \RMMrevise{the critical layers} $y=\pm 0.79$\sout{, which corresponds to the critical layers} for the TS wavespeed of $c_r\approx 0.37$. Comparison with Figs.~\ref{fig:Cxx_comparison}a and \ref{fig:Cxx_comparison}b shows a strong similarity between the eigenmode and the tilted sheetlike structures that are the hallmark of EIT, with the resemblance between the TS mode and the $(k_x L_x/2\pi,k_z L_z/2\pi)=(2,0)$ structure from the DNS\sout{ shown} in Fig.~\ref{fig:Cxx_comparison}b 
being particularly striking.  
{Specifically, note that for the TS mode, Fig.~\ref{fig:Cxx_comparison}c, $v'$ and $\alpha_{xx}'$ are even and odd, respectively, with respect to $y=0$, while in Fig.~\ref{fig:Cxx_comparison}b and the corresponding results at higher wavenumbers, these symmetries hold to a good approximation.}

Despite the fact that the TS mode ultimately {decays},  the non-normal character of the linearized Navier-Stokes operator can lead to significant disturbance growth at short times or significant amplification of harmonic-in-time disturbances \cite{Schmid07}. 
\MDGrevise{Thus it} \sout{It }is therefore possible for small disturbances to be sufficiently amplified \MDGrevise{that }\sout{for }nonlinear effects \sout{to }become significant.
We now quantify this amplification by computing the largest singular value \RMMrevise{$\sigma_1$} of the resolvent operator\sout{ of the viscoelastic governing equations linearized around the laminar state}.\sout{ This is the transfer function between input disturbances and the system response at a given real frequency $\omega$.} 
Figure~\ref{fig:Combine_LA}b shows \sout{the}\RMMrevise{results}\sout{largest singular value $\sigma_1$} for $\Wi=0$ and $\Wi=20$ in the same range of \MDGrevise{(real)} wavespeeds {$c=\omega/k_x$} depicted in Figure~\ref{fig:Combine_LA}a. \MDGrevise{The amplification increases dramatically with}\sout{There is a dramatic increase in amplification with increasing} $\Wi$, with the values at $\Wi=20$ being \RMMrevise{$\sim\!10^2$ times}\sout{approximately two orders of magnitude larger than} those for $\Wi=0$; this is {consistent with} the drastic increase in the conformation tensor disturbance amplitude already discussed for the TS mode. 
In both cases, the maximum amplification occurs for $c\approx 0.37$, which coincides with the wavespeed for the TS mode, indicating that the most-amplified disturbance is closely linked to the TS wave. Figure~\ref{fig:Cxx_comparison}d shows the leading resolvent mode, which is indeed almost identical to the TS eigenmode in Figure~\ref{fig:Cxx_comparison}c. This result provides additional strong evidence that the structures observed in EIT are closely related to those in viscoelasticity-modified TS waves.

{It was recently shown that viscoelastic pipe flow of an Oldroyd-B fluid ($b\rightarrow\infty$) can be linearly unstable to\sout{ a} center-localized modes with wavespeed $c_r\approx 1$~\cite{garg2018}. We estimate that for the present parameter values, this mode only becomes relevant for very high $\Wi$. Furthermore, center-localized structures are not observed in the simulations of EIT, so we do not consider them relevant here.}

\paragraph{Self-sustained viscoelastic Tollmien-Schlichting waves:}

    \begin{figure}[]
    \begin{center}
	\includegraphics[scale=0.08]{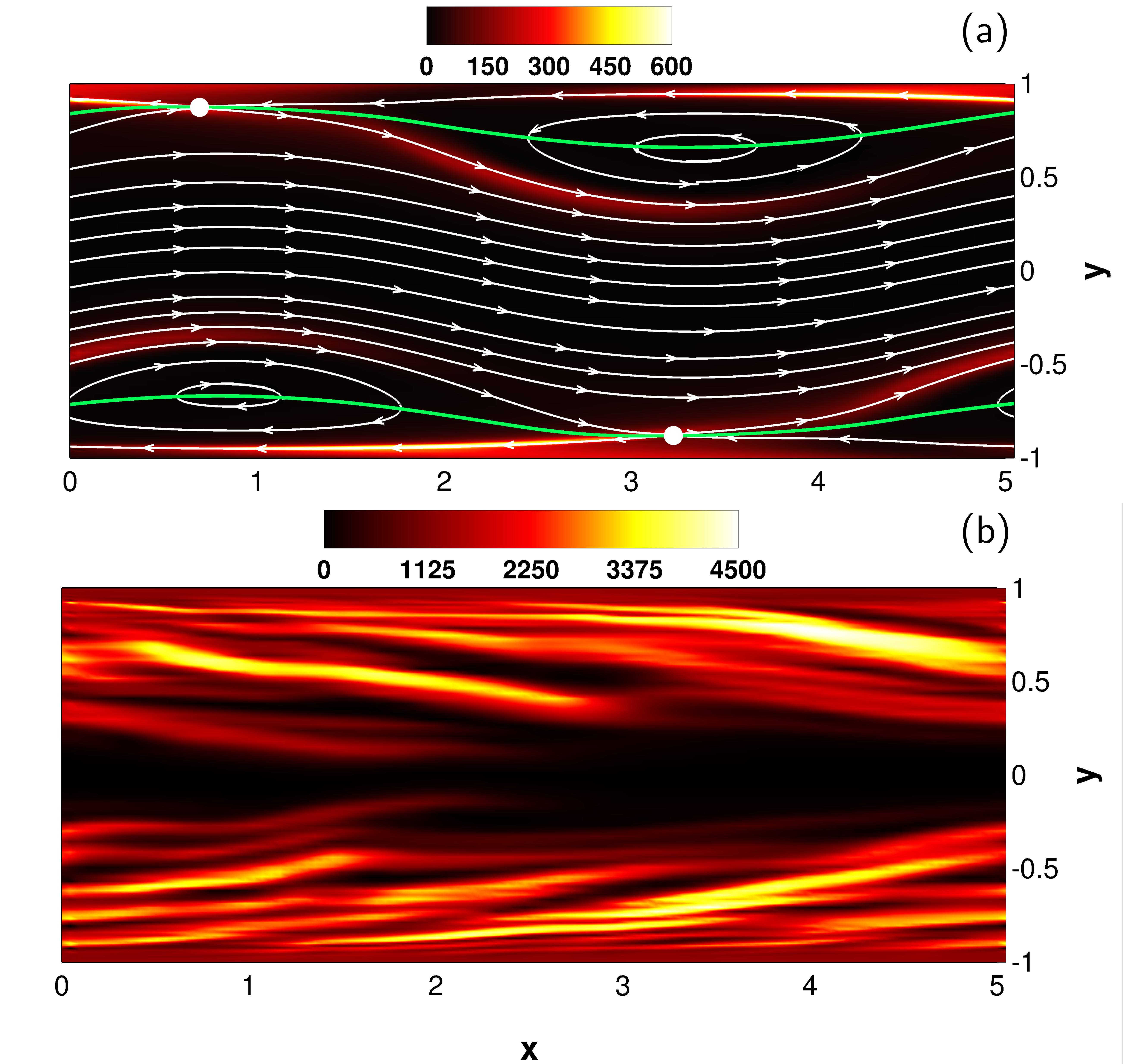}
	\caption[]{(a) Structure of nonlinear self-sustaining TS wave at $\Rey = 3000, \Wi = 3$. White streamlines{, shown in a reference frame moving with the wavespeed $c=0.39$,} are superimposed on color contours of $\alpha_{xx}$.  Green lines indicate the instantaneous critical layer positions, and white dots indicate the locations of hyperbolic stagnation points. (b) Snapshot of $\alpha_{xx}$ contours from 2D EIT at $\Rey = 3000$, $\Wi = 15$.
}
 	\label{fig:Streamline_7}
    \end{center}
    \end{figure}

 Here we elaborate on the potential connection between TS-like structure and EIT, presenting results for nonlinear viscoelastic TS waves, i.e.~self-sustained traveling wave solutions of the full nonlinear governing equations, illustrating the role of the critical-layer kinematics in generating localized sheetlike regions of high polymer stretching like those observed in EIT.

The strong peak in the EIT spectrum seen in Figure \ref{fig:fminflam} corresponds to a wavelength of $5$, so here we report computations of nonlinear TS wave in a 2D domain with this length. The upper branch of this solution family is linearly stable in 2D at $\Rey = 3000$ \cite{Jimenez:1990cn,Mellibovsky:2015ep,herbert79} and easily captured \MDGrevise{with DNS} using the linear TS mode as the initial condition. In Newtonian flow, the solution family exists at this wavelength down to $\Rey \approx 2800$. {We continue the} Newtonian solution at $\Rey = 3000$ to the parameters of interest ($\beta = 0.97$ and $b = 6400$) at $\Wi = 0.1$, then \MDGrevise{increase} $\Wi$ \sout{is increased }to study the effect of viscoelasticity. Hameduddin et al.~\cite{hameduddin2019} have 
computed nonlinear viscoelastic TS waves in the \sout{supercritical}\MDGrevise{regime} $\Rey>5772$\sout{ regime} and noted the role the critical layer plays in polymer stretching at high $\Wi$, but have not reported the observations described below. 

On increasing $\Wi$, the self-sustained nonlinear viscoelastic TS wave at $\Rey=3000$ develops \sout{a structure of} sheets of high polymer stretch \sout{bearing resemblance to }\MDGrevise{resembling} near wall structures seen at EIT. Figure \ref{fig:Streamline_7} illustrates this \sout{observation}\MDGrevise{point} with a plot of $\alpha_{xx}$ at $\Wi=3$. The source of this stretching is closely tied to the critical-layer structure {of} the TS wave velocity field.  
Critical layers have long been-known to exhibit a so-called Kelvin cat's-eye \MDGrevise{ streamline }structure \cite{DrazinReid} -- indeed, the velocity fields \MDGrevise{for the flows shown in}\sout{ of both the linear TS mode and the dominant resolvent mode reported in} Figures \ref{fig:Cxx_comparison}c and \ref{fig:Cxx_comparison}d display this feature.
With regard to viscoelasticity, the cat's-eye structure is important because it contains hyperbolic stagnation points: polymers are strongly stretched as they approach such points and leave along their unstable manifolds. This phenomenon is clearly seen in Figure \ref{fig:Streamline_7}a; shown in white are streamlines in the reference frame traveling with the speed of the wave $c=0.39$, and in green is the instantaneous critical-layer position, i.e.~where $v_x=c$. A hyperbolic stagnation point (white dot) exists at $x = 3.22, y = -0.87$.  The high polymer stretching follows the streamlines along the unstable directions associated with this point, giving rise to an arched sheetlike structure. By symmetry, identical structures  exist in the top half of the channel.  For comparison, Figure \ref{fig:Streamline_7}b shows $\alpha_{xx}$ for 2D EIT at $\Rey = 3000, \Wi = 15$. This takes the form of tilted sheets of high polymer stretch starting out at locations close to the walls, and in fact reasonably close to the positions $y=\pm 0.87$ of the stagnation points in the nonlinear TS wave at $\Wi=3$. This similarity in structures suggests a role for TS wave-like critical-layer mechanisms at EIT.
Indeed, these results suggest that the nonlinear TS wave solution branch may be directly connected in parameter space to EIT. We do not find this to be the case at $\Rey=3000$\MDGrevise{;}\sout{, where} the TS branch loses existence above $\Wi\approx 4$ and the EIT branch loses existence below $\Wi\approx 13$. {Nevertheless, when using the EIT result at $\Wi=13$ as the initial condition for a simulation at $\Wi=12$,  EIT persists transiently for hundreds of time units and the last remaining structure observed as the flow decays to laminar closely resembles Figs.~2b-d.}

\paragraph{Conclusion}

{Elastoinertial turbulence at low $\Rey$ has strongly localized stress fluctuations, suggesting the importance of critical-layer mechanisms in its origin. These fluctuations strongly resemble the most slowly decaying structures from linear stability analysis, as well as the most strongly amplified disturbances as determined by resolvent analysis of the linearized equations. Furthermore, the Kelvin cat's eye \sout{structure}\MDGrevise{kinematics} found in the critical-layer region of self-sustained nonlinear TS waves\sout{ has kinematics that} generate sheetlike structures in the stress field that resemble those observed in EIT. Taken together, these results suggest that, at least in the parameter range considered here, the bypass transition leading to EIT is mediated by nonlinear amplification and self-sustenance of perturbations that generate TS-wave-like flow structures.}



This work was supported by (UW) NSF through grant CBET-1510291, AFOSR through grants FA9550-15-1-0062 and FA9550-18-1-0174, and ONR through grants N00014-18-1-2865 and (Caltech) N00014-17-1-3022.

A.S. and R.M.M. contributed equally to this work.


\begin{thebibliography}{0}%
\makeatletter
\providecommand \@ifxundefined [1]{%
 \@ifx{#1\undefined}
}%
\providecommand \@ifnum [1]{%
 \ifnum #1\expandafter \@firstoftwo
 \else \expandafter \@secondoftwo
 \fi
}%
\providecommand \@ifx [1]{%
 \ifx #1\expandafter \@firstoftwo
 \else \expandafter \@secondoftwo
 \fi
}%
\providecommand \natexlab [1]{#1}%
\providecommand \enquote  [1]{``#1''}%
\providecommand \bibnamefont  [1]{#1}%
\providecommand \bibfnamefont [1]{#1}%
\providecommand \citenamefont [1]{#1}%
\providecommand \href@noop [0]{\@secondoftwo}%
\providecommand \href [0]{\begingroup \@sanitize@url \@href}%
\providecommand \@href[1]{\@@startlink{#1}\@@href}%
\providecommand \@@href[1]{\endgroup#1\@@endlink}%
\providecommand \@sanitize@url [0]{\catcode `\\12\catcode `\$12\catcode
  `\&12\catcode `\#12\catcode `\^12\catcode `\_12\catcode `\%12\relax}%
\providecommand \@@startlink[1]{}%
\providecommand \@@endlink[0]{}%
\providecommand \url  [0]{\begingroup\@sanitize@url \@url }%
\providecommand \@url [1]{\endgroup\@href {#1}{\urlprefix }}%
\providecommand \urlprefix  [0]{URL }%
\providecommand \Eprint [0]{\href }%
\providecommand \doibase [0]{http://dx.doi.org/}%
\providecommand \selectlanguage [0]{\@gobble}%
\providecommand \bibinfo  [0]{\@secondoftwo}%
\providecommand \bibfield  [0]{\@secondoftwo}%
\providecommand \translation [1]{[#1]}%
\providecommand \BibitemOpen [0]{}%
\providecommand \bibitemStop [0]{}%
\providecommand \bibitemNoStop [0]{.\EOS\space}%
\providecommand \EOS [0]{\spacefactor3000\relax}%
\providecommand \BibitemShut  [1]{\csname bibitem#1\endcsname}%
\let\auto@bib@innerbib\@empty
\end{thebibliography}%


\begin{thebibliography}{23}%
\makeatletter
\providecommand \@ifxundefined [1]{%
 \@ifx{#1\undefined}
}%
\providecommand \@ifnum [1]{%
 \ifnum #1\expandafter \@firstoftwo
 \else \expandafter \@secondoftwo
 \fi
}%
\providecommand \@ifx [1]{%
 \ifx #1\expandafter \@firstoftwo
 \else \expandafter \@secondoftwo
 \fi
}%
\providecommand \natexlab [1]{#1}%
\providecommand \enquote  [1]{``#1''}%
\providecommand \bibnamefont  [1]{#1}%
\providecommand \bibfnamefont [1]{#1}%
\providecommand \citenamefont [1]{#1}%
\providecommand \href@noop [0]{\@secondoftwo}%
\providecommand \href [0]{\begingroup \@sanitize@url \@href}%
\providecommand \@href[1]{\@@startlink{#1}\@@href}%
\providecommand \@@href[1]{\endgroup#1\@@endlink}%
\providecommand \@sanitize@url [0]{\catcode `\\12\catcode `\$12\catcode
  `\&12\catcode `\#12\catcode `\^12\catcode `\_12\catcode `\%12\relax}%
\providecommand \@@startlink[1]{}%
\providecommand \@@endlink[0]{}%
\providecommand \url  [0]{\begingroup\@sanitize@url \@url }%
\providecommand \@url [1]{\endgroup\@href {#1}{\urlprefix }}%
\providecommand \urlprefix  [0]{URL }%
\providecommand \Eprint [0]{\href }%
\providecommand \doibase [0]{http://dx.doi.org/}%
\providecommand \selectlanguage [0]{\@gobble}%
\providecommand \bibinfo  [0]{\@secondoftwo}%
\providecommand \bibfield  [0]{\@secondoftwo}%
\providecommand \translation [1]{[#1]}%
\providecommand \BibitemOpen [0]{}%
\providecommand \bibitemStop [0]{}%
\providecommand \bibitemNoStop [0]{.\EOS\space}%
\providecommand \EOS [0]{\spacefactor3000\relax}%
\providecommand \BibitemShut  [1]{\csname bibitem#1\endcsname}%
\let\auto@bib@innerbib\@empty
\bibitem [{\citenamefont {Virk}\ \emph {et~al.}(1970)\citenamefont {Virk},
  \citenamefont {Mickley},\ and\ \citenamefont {Smith}}]{Virk:1970uo}%
  \BibitemOpen
  \bibfield  {author} {\bibinfo {author} {\bibfnamefont {P.~S.}\ \bibnamefont
  {Virk}}, \bibinfo {author} {\bibfnamefont {H.~S.}\ \bibnamefont {Mickley}}, \
  and\ \bibinfo {author} {\bibfnamefont {K.~A.}\ \bibnamefont {Smith}},\
  }\href@noop {} {\bibfield  {journal} {\bibinfo  {journal} {J. Appl. Mech.}\
  }\textbf {\bibinfo {volume} {37}},\ \bibinfo {pages} {488} (\bibinfo {year}
  {1970})}\BibitemShut {NoStop}%
\bibitem [{\citenamefont {Forame}\ \emph {et~al.}(1972)\citenamefont {Forame},
  \citenamefont {Hansen},\ and\ \citenamefont {Little}}]{Forame:1972ti}%
  \BibitemOpen
  \bibfield  {author} {\bibinfo {author} {\bibfnamefont {P.~C.}\ \bibnamefont
  {Forame}}, \bibinfo {author} {\bibfnamefont {R.~J.}\ \bibnamefont {Hansen}},
  \ and\ \bibinfo {author} {\bibfnamefont {R.~C.}\ \bibnamefont {Little}},\
  }\href@noop {} {\bibfield  {journal} {\bibinfo  {journal} {AIChE Journal}\
  }\textbf {\bibinfo {volume} {18}},\ \bibinfo {pages} {213} (\bibinfo {year}
  {1972})}\BibitemShut {NoStop}%
\bibitem [{\citenamefont {Hoyt}(1977)}]{Hoyt:1977vj}%
  \BibitemOpen
  \bibfield  {author} {\bibinfo {author} {\bibfnamefont {J.~W.}\ \bibnamefont
  {Hoyt}},\ }\href@noop {} {\bibfield  {journal} {\bibinfo  {journal} {Nature}\
  }\textbf {\bibinfo {volume} {270}},\ \bibinfo {pages} {508} (\bibinfo {year}
  {1977})}\BibitemShut {NoStop}%
\bibitem [{\citenamefont {Choueiri}\ \emph {et~al.}(2018)\citenamefont
  {Choueiri}, \citenamefont {Lopez},\ and\ \citenamefont
  {Hof}}]{Choueiri:2018it}%
  \BibitemOpen
  \bibfield  {author} {\bibinfo {author} {\bibfnamefont {G.~H.}\ \bibnamefont
  {Choueiri}}, \bibinfo {author} {\bibfnamefont {J.~M.}\ \bibnamefont {Lopez}},
  \ and\ \bibinfo {author} {\bibfnamefont {B.}~\bibnamefont {Hof}},\
  }\href@noop {} {\bibfield  {journal} {\bibinfo  {journal} {Phys. Rev. Lett.}\
  }\textbf {\bibinfo {volume} {120}},\ \bibinfo {pages} {124501} (\bibinfo
  {year} {2018})}\BibitemShut {NoStop}%
\bibitem [{\citenamefont {Chandra}\ \emph {et~al.}(2018)\citenamefont
  {Chandra}, \citenamefont {Shankar},\ and\ \citenamefont
  {Das}}]{Chandra:2018bq}%
  \BibitemOpen
  \bibfield  {author} {\bibinfo {author} {\bibfnamefont {B.}~\bibnamefont
  {Chandra}}, \bibinfo {author} {\bibfnamefont {V.}~\bibnamefont {Shankar}}, \
  and\ \bibinfo {author} {\bibfnamefont {D.}~\bibnamefont {Das}},\ }\href@noop
  {} {\bibfield  {journal} {\bibinfo  {journal} {J. Fluid Mech.}\ }\textbf
  {\bibinfo {volume} {844}},\ \bibinfo {pages} {1052} (\bibinfo {year}
  {2018})}\BibitemShut {NoStop}%
\bibitem [{\citenamefont {Samanta}\ \emph {et~al.}(2013)\citenamefont
  {Samanta}, \citenamefont {Dubief}, \citenamefont {Holzner}, \citenamefont
  {Sch{\"a}fer}, \citenamefont {Morozov}, \citenamefont {Wagner},\ and\
  \citenamefont {Hof}}]{Samanta:2013el}%
  \BibitemOpen
  \bibfield  {author} {\bibinfo {author} {\bibfnamefont {D.}~\bibnamefont
  {Samanta}}, \bibinfo {author} {\bibfnamefont {Y.}~\bibnamefont {Dubief}},
  \bibinfo {author} {\bibfnamefont {M.}~\bibnamefont {Holzner}}, \bibinfo
  {author} {\bibfnamefont {C.}~\bibnamefont {Sch{\"a}fer}}, \bibinfo {author}
  {\bibfnamefont {A.~N.}\ \bibnamefont {Morozov}}, \bibinfo {author}
  {\bibfnamefont {C.}~\bibnamefont {Wagner}}, \ and\ \bibinfo {author}
  {\bibfnamefont {B.}~\bibnamefont {Hof}},\ }\href@noop {} {\bibfield
  {journal} {\bibinfo  {journal} {Proc. Nat. Acad. Sci.}\ }\textbf {\bibinfo
  {volume} {110}},\ \bibinfo {pages} {10557} (\bibinfo {year}
  {2013})}\BibitemShut {NoStop}%
\bibitem [{\citenamefont {Dubief}\ \emph {et~al.}(2013)\citenamefont {Dubief},
  \citenamefont {Terrapon},\ and\ \citenamefont {Soria}}]{Dubief:2013hh}%
  \BibitemOpen
  \bibfield  {author} {\bibinfo {author} {\bibfnamefont {Y.}~\bibnamefont
  {Dubief}}, \bibinfo {author} {\bibfnamefont {V.~E.}\ \bibnamefont
  {Terrapon}}, \ and\ \bibinfo {author} {\bibfnamefont {J.}~\bibnamefont
  {Soria}},\ }\href@noop {} {\bibfield  {journal} {\bibinfo  {journal} {Phys.
  Fluids}\ }\textbf {\bibinfo {volume} {25}},\ \bibinfo {pages} {110817}
  (\bibinfo {year} {2013})}\BibitemShut {NoStop}%
\bibitem [{\citenamefont {Sid}\ \emph {et~al.}(2018)\citenamefont {Sid},
  \citenamefont {Terrapon},\ and\ \citenamefont {Dubief}}]{Sid:2018gh}%
  \BibitemOpen
  \bibfield  {author} {\bibinfo {author} {\bibfnamefont {S.}~\bibnamefont
  {Sid}}, \bibinfo {author} {\bibfnamefont {V.~E.}\ \bibnamefont {Terrapon}}, \
  and\ \bibinfo {author} {\bibfnamefont {Y.}~\bibnamefont {Dubief}},\
  }\href@noop {} {\bibfield  {journal} {\bibinfo  {journal} {Phys. Rev.
  Fluids}\ }\textbf {\bibinfo {volume} {3}},\ \bibinfo {pages} {011301}
  (\bibinfo {year} {2018})}\BibitemShut {NoStop}%
\bibitem [{\citenamefont {Kurganov}\ and\ \citenamefont
  {Tadmor}(2000)}]{kurganov2000new}%
  \BibitemOpen
  \bibfield  {author} {\bibinfo {author} {\bibfnamefont {A.}~\bibnamefont
  {Kurganov}}\ and\ \bibinfo {author} {\bibfnamefont {E.}~\bibnamefont
  {Tadmor}},\ }\href@noop {} {\bibfield  {journal} {\bibinfo  {journal} {J.
  Comput. Phys.}\ }\textbf {\bibinfo {volume} {160}},\ \bibinfo {pages} {241}
  (\bibinfo {year} {2000})}\BibitemShut {NoStop}%
\bibitem [{\citenamefont {Vaithianathan}\ \emph {et~al.}(2006)\citenamefont
  {Vaithianathan}, \citenamefont {Robert}, \citenamefont {Brasseur},\ and\
  \citenamefont {Collins}}]{Vaithianathan:2006dy}%
  \BibitemOpen
  \bibfield  {author} {\bibinfo {author} {\bibfnamefont {T.}~\bibnamefont
  {Vaithianathan}}, \bibinfo {author} {\bibfnamefont {A.}~\bibnamefont
  {Robert}}, \bibinfo {author} {\bibfnamefont {J.~G.}\ \bibnamefont
  {Brasseur}}, \ and\ \bibinfo {author} {\bibfnamefont {L.~R.}\ \bibnamefont
  {Collins}},\ }\href@noop {} {\bibfield  {journal} {\bibinfo  {journal} {J.
  Non-Newtonian Fluid Mech.}\ }\textbf {\bibinfo {volume} {140}},\ \bibinfo
  {pages} {3} (\bibinfo {year} {2006})}\BibitemShut {NoStop}%
\bibitem [{\citenamefont {Dallas}\ \emph {et~al.}(2010)\citenamefont {Dallas},
  \citenamefont {Vassilicos},\ and\ \citenamefont {Hewitt}}]{Dallas:2010gu}%
  \BibitemOpen
  \bibfield  {author} {\bibinfo {author} {\bibfnamefont {V.}~\bibnamefont
  {Dallas}}, \bibinfo {author} {\bibfnamefont {J.}~\bibnamefont {Vassilicos}},
  \ and\ \bibinfo {author} {\bibfnamefont {G.}~\bibnamefont {Hewitt}},\
  }\href@noop {} {\bibfield  {journal} {\bibinfo  {journal} {Phys. Rev. E}\
  }\textbf {\bibinfo {volume} {82}},\ \bibinfo {pages} {066303} (\bibinfo
  {year} {2010})}\BibitemShut {NoStop}%
\bibitem [{\citenamefont {Hameduddin}\ \emph {et~al.}(2019)\citenamefont
  {Hameduddin}, \citenamefont {Gayme},\ and\ \citenamefont
  {Zaki}}]{hameduddin2019}%
  \BibitemOpen
  \bibfield  {author} {\bibinfo {author} {\bibfnamefont {I.}~\bibnamefont
  {Hameduddin}}, \bibinfo {author} {\bibfnamefont {D.~F.}\ \bibnamefont
  {Gayme}}, \ and\ \bibinfo {author} {\bibfnamefont {T.~A.}\ \bibnamefont
  {Zaki}},\ }\href@noop {} {\bibfield  {journal} {\bibinfo  {journal} {J. Fluid
  Mech.}\ }\textbf {\bibinfo {volume} {858}},\ \bibinfo {pages} {377} (\bibinfo
  {year} {2019})}\BibitemShut {NoStop}%
\bibitem [{\citenamefont {Schmid}(2007)}]{Schmid07}%
  \BibitemOpen
  \bibfield  {author} {\bibinfo {author} {\bibfnamefont {P.~J.}\ \bibnamefont
  {Schmid}},\ }\href@noop {} {\bibfield  {journal} {\bibinfo  {journal} {Annu.
  Rev. Fluid Mech.}\ }\textbf {\bibinfo {volume} {39}},\ \bibinfo {pages} {129}
  (\bibinfo {year} {2007})}\BibitemShut {NoStop}%
\bibitem [{\citenamefont {McKeon}\ and\ \citenamefont
  {Sharma}(2010)}]{mckeon2010}%
  \BibitemOpen
  \bibfield  {author} {\bibinfo {author} {\bibfnamefont {B.~J.}~\bibnamefont
  {McKeon}}\ and\ \bibinfo {author} {\bibfnamefont {A.~S.}~\bibnamefont
  {Sharma}},\ }\href@noop {} {\bibfield  {journal} {\bibinfo  {journal} {J.
  Fluid Mech.}\ }\textbf {\bibinfo {volume} {658}},\ \bibinfo {pages} {336}
  (\bibinfo {year} {2010})}\BibitemShut {NoStop}%
\bibitem [{\citenamefont {Page}\ and\ \citenamefont
  {Zaki}(2015)}]{Page:2015es}%
  \BibitemOpen
  \bibfield  {author} {\bibinfo {author} {\bibfnamefont {J.}~\bibnamefont
  {Page}}\ and\ \bibinfo {author} {\bibfnamefont {T.~A.}\ \bibnamefont
  {Zaki}},\ }\href@noop {} {\bibfield  {journal} {\bibinfo  {journal} {J. Fluid
  Mech.}\ }\textbf {\bibinfo {volume} {777}},\ \bibinfo {pages} {327} (\bibinfo
  {year} {2015})}\BibitemShut {NoStop}%
\bibitem [{\citenamefont {Lee}\ and\ \citenamefont {Zaki}(2017)}]{Lee:2017fb}%
  \BibitemOpen
  \bibfield  {author} {\bibinfo {author} {\bibfnamefont {S.~J.}\ \bibnamefont
  {Lee}}\ and\ \bibinfo {author} {\bibfnamefont {T.~A.}\ \bibnamefont {Zaki}},\
  }\href@noop {} {\bibfield  {journal} {\bibinfo  {journal} {J. Fluid Mech.}\
  }\textbf {\bibinfo {volume} {820}},\ \bibinfo {pages} {232} (\bibinfo {year}
  {2017})}\BibitemShut {NoStop}%
\bibitem [{\citenamefont {Haward}\ \emph {et~al.}(2018)\citenamefont {Haward},
  \citenamefont {Page}, \citenamefont {Zaki},\ and\ \citenamefont
  {Shen}}]{Haward:2018hs}%
  \BibitemOpen
  \bibfield  {author} {\bibinfo {author} {\bibfnamefont {S.~J.}\ \bibnamefont
  {Haward}}, \bibinfo {author} {\bibfnamefont {J.}~\bibnamefont {Page}},
  \bibinfo {author} {\bibfnamefont {T.~A.}\ \bibnamefont {Zaki}}, \ and\
  \bibinfo {author} {\bibfnamefont {A.~Q.}\ \bibnamefont {Shen}},\ }\href@noop
  {} {\bibfield  {journal} {\bibinfo  {journal} {Phys. Rev. Fluids}\ }\textbf
  {\bibinfo {volume} {3}},\ \bibinfo {pages} {091302} (\bibinfo {year}
  {2018})}\BibitemShut {NoStop}%
\bibitem [{\citenamefont {Drazin}\ and\ \citenamefont
  {Reid}(2004)}]{DrazinReid}%
  \BibitemOpen
  \bibfield  {author} {\bibinfo {author} {\bibfnamefont {P.~G.}\ \bibnamefont
  {Drazin}}\ and\ \bibinfo {author} {\bibfnamefont {W.~H.}\ \bibnamefont
  {Reid}},\ }\href@noop {} {\emph {\bibinfo {title} {Hydrodynamic
  {S}tability}}},\ \bibinfo {edition} {2nd}\ ed.,\ Cambridge Mathematical
  Libraries\ (\bibinfo  {publisher} {Cambridge University Press},\ \bibinfo
  {year} {2004})\BibitemShut {NoStop}%
\bibitem [{\citenamefont {Zhang}\ \emph {et~al.}(2013)\citenamefont {Zhang},
  \citenamefont {Lashgari}, \citenamefont {Zaki},\ and\ \citenamefont
  {Brandt}}]{zhang2013}%
  \BibitemOpen
  \bibfield  {author} {\bibinfo {author} {\bibfnamefont {M.}~\bibnamefont
  {Zhang}}, \bibinfo {author} {\bibfnamefont {I.}~\bibnamefont {Lashgari}},
  \bibinfo {author} {\bibfnamefont {T.~A.}\ \bibnamefont {Zaki}}, \ and\
  \bibinfo {author} {\bibfnamefont {L.}~\bibnamefont {Brandt}},\ }\href@noop {}
  {\bibfield  {journal} {\bibinfo  {journal} {J. Fluid Mech.}\ }\textbf
  {\bibinfo {volume} {737}},\ \bibinfo {pages} {249} (\bibinfo {year}
  {2013})}\BibitemShut {NoStop}%
\bibitem [{\citenamefont {Garg}\ \emph {et~al.}(2018)\citenamefont {Garg},
  \citenamefont {Chaudhary}, \citenamefont {Khalid}, \citenamefont {Shankar},\
  and\ \citenamefont {Subramanian}}]{garg2018}%
  \BibitemOpen
  \bibfield  {author} {\bibinfo {author} {\bibfnamefont {P.}~\bibnamefont
  {Garg}}, \bibinfo {author} {\bibfnamefont {I.}~\bibnamefont {Chaudhary}},
  \bibinfo {author} {\bibfnamefont {M.}~\bibnamefont {Khalid}}, \bibinfo
  {author} {\bibfnamefont {V.}~\bibnamefont {Shankar}}, \ and\ \bibinfo
  {author} {\bibfnamefont {G.}~\bibnamefont {Subramanian}},\ }\href@noop {}
  {\bibfield  {journal} {\bibinfo  {journal} {Phys. Rev. Lett.}\ }\textbf
  {\bibinfo {volume} {121}},\ \bibinfo {pages} {024502} (\bibinfo {year}
  {2018})}\BibitemShut {NoStop}%
\bibitem [{\citenamefont {Jim{\'e}nez}(1990)}]{Jimenez:1990cn}%
  \BibitemOpen
  \bibfield  {author} {\bibinfo {author} {\bibfnamefont {J.}~\bibnamefont
  {Jim{\'e}nez}},\ }\href@noop {} {\bibfield  {journal} {\bibinfo  {journal}
  {J. Fluid Mech.}\ }\textbf {\bibinfo {volume} {218}},\ \bibinfo {pages} {265}
  (\bibinfo {year} {1990})}\BibitemShut {NoStop}%
\bibitem [{\citenamefont {Mellibovsky}\ and\ \citenamefont
  {Meseguer}(2015)}]{Mellibovsky:2015ep}%
  \BibitemOpen
  \bibfield  {author} {\bibinfo {author} {\bibfnamefont {F.}~\bibnamefont
  {Mellibovsky}}\ and\ \bibinfo {author} {\bibfnamefont {A.}~\bibnamefont
  {Meseguer}},\ }\href@noop {} {\bibfield  {journal} {\bibinfo  {journal} {J.
  Fluid Mech.}\ }\textbf {\bibinfo {volume} {779}},\ \bibinfo {pages} {R1}
  (\bibinfo {year} {2015})}\BibitemShut {NoStop}%
\bibitem [{\citenamefont {Herbert}(1979)}]{herbert79}%
  \BibitemOpen
  \bibfield  {author} {\bibinfo {author} {\bibfnamefont {T.}~\bibnamefont
  {Herbert}},\ }in\ \href@noop {} {\emph {\bibinfo {booktitle} {Proceedings of
  the Fifth International Conference on Numerical Methods in Fluid Dynamics
  June 28--July 2, 1976 Twente University, Enschede}}}\ (\bibinfo
  {organization} {Springer},\ \bibinfo {year} {1979})\ pp.\ \bibinfo {pages}
  {235--240}\BibitemShut {NoStop}%
\end{thebibliography}

%

\end{document}